\documentclass[10pt]{article}
\usepackage[english]{babel}
\usepackage{amsmath,amsfonts,amssymb,bm}\interdisplaylinepenalty=2500
\ifx\pdfoutput\undefined
  \usepackage{graphicx,color}   
  \usepackage[nativepdf,backref,bookmarks]{hyperref} 
  \usepackage{rotating}
\else
  \usepackage[backref,bookmarks]{hyperref}  
  \usepackage[pdftex]{graphicx,color}     
  \usepackage[pdftex]{rotating}              
\fi
\ifx\undefined\degrees\def\degrees{\ensuremath{^{\circ}}}\fi
\ifx\undefined\celsius\def\celsius{\ensuremath{^{\circ}\mathrm{C}}}\fi
\ifx\undefined\unit\def\unit#1{\ensuremath{\mathrm{\,#1}}}\fi
\ifx\undefined\micro\def\micro{\ensuremath{\mu}}\fi
\ifx\undefined\sups\def\sups#1{\ensuremath{^{\mathrm{#1}}}}\fi
\ifx\undefined\subs\def\subs#1{\ensuremath{_{\mathrm{#1}}}}\fi
\ifx\undefined\ohm\def\ohm{\ensuremath{\mathrm{\Omega}}}\fi

\begin{document}
\raggedbottom

\title{Low flicker-noise DC amplifier for 50~\ohm\ sources}
\author{Enrico Rubiola\thanks{%
    Universit\'e Henri Poincar\'e, Nancy, France,
      \texttt{www.rubiola.org}, e-mail \texttt{enrico@rubiola.org} }
   \and Franck Lardet-Vieudrin\thanks{%
     Dept.~\textsc{lpmo}, \textsc{femto-st} Besan\c{c}on, France,
     e-mail \texttt{lardet@lpmo.edu}}}
\date{\small Cite this article as:\\
  E. Rubiola, F. Lardet-Vieudrin, ``Low flicker-noise amplifier for 50~\ohm\ sources'',  
  \emph{Review of Scientific Instruments} vol.~75 no.~5 pp.1323--1326, May 2004}

\maketitle

\begin{abstract}
  This article analyzes the design of a low-noise amplifier intended
  as the input front-end for the measurement of the low-frequency
  components (below 10 Hz) of a 50 \ohm\ source.  Low residual flicker
  is the main desired performance.  This feature can only be
  appreciated if white noise is sufficiently low, and if an
  appropriate design ensures dc stability.  An optimal solution is
  proposed, in which the low-noise and dc-stability features are
  achieved at a reasonable complexity.  Gain is accurate to more than
  100 kHz, which makes the amplifier an appealing external front-end
  for fast Fourier transform (FFT) analyzers.
\end{abstract}

\tableofcontents

\section{Introduction}\label{sec:intro}
%
Often the experimentalist needs a low-noise preamplifier for the
analysis of low-frequency components (below 10 Hz) from a 50 \ohm\ 
source.  The desired amplifier chiefly exhibits low residual flicker
and high thermal stability, besides low white noise.  Thermal
stability without need for temperature control is a desirable feature.
In fact the problem with temperature control, worse than complexity,
is that in a nonstabilized environment thermal gradients fluctuate,
and in turn low-frequency noise is taken in.  A low-noise amplifier
may be regarded as an old subject, nonetheless innovation in analysis
methods and in available parts provides insight and new design.  The
application we initially had in mind is the postdetection preamplifier
for phase noise measurements~\cite{rubiola02rsi}.  Yet, there resulted
a versatile general-purpose scheme useful in experimental electronics
and physics.

\section{Design Strategy}\label{sec:strategy}
%
The choice of the input stage determines the success of a precision
amplifier.  This issue involves the choice of appropriate devices and
of the topology.

Available low-noise devices are the junction field-effect transistor
(JFET) and the bipolar transistor (BJT), either as part of an
operational amplifier or as a stand-alone component.  The white noise
of these devices is well
understood~\cite{van.der.ziel:noise-ssdc,van.der.ziel:fluctuations,netzer81pieee,erdi81jssc}.
Conversely, flicker noise is still elusive and relies upon models, the
most accredited of which are due to McWhorter~\cite{mcwhorter57} and
Hooge~\cite{hooge69pla}, or on smart narrow-domain analyses, like~\cite{green85jpd-1,green85jpd-2,jamaldeen99jap}, rather than on a unified theory.  Even worse,
aging and thermal drift chiefly depend on proprietary technologies,
thus scientific literature ends up to be of scarce usefulness.  The
JFET is appealing because of the inherently low white noise.  The
noise temperature can be as low as a fraction of a degree Kelvin.
Unfortunately, the low noise of the JFET derives from low input
current, hence a high input resistance (some M\ohm) is necessary.  The
JFET noise voltage is hardly lower than 5 \unit{nV/\sqrt{Hz}}, some
five to six times higher than the thermal noise of a 50 \ohm\ resistor
($\sqrt{4kTR}=0.89$ \unit{nV/\sqrt{Hz}}).  The JFET is therefore
discarded in favor of the BJT\@.

A feedback scheme, in which the gain is determined by a resistive
network, is necessary for gain accuracy and flatness over frequency.
Besides the well known differential stage, a single-transistor
configuration is possible (Ref.~\cite{motchenbacher:low-noise:1ed},
page 123), in which the input is connected to the base and the
feedback to the emitter.  This configuration was popular in early
audio hi-fi amplifiers.  The advantage of the single-transistor scheme
is that noise power is half the noise of a differential stage.  On the
other hand, in a dc-coupled circuit thermal effects are difficult to
compensate without reintroducing noise, while thermal compensation of
the differential stage is guaranteed by the symmetry of the
base-emitter junctions.  Hence we opt for the differential pair.

\begin{table}
\begin{sideways}
\begin{minipage}{0.88\textheight}
\caption{\label{tab:opa}%
  \vrule width0pt height2.5ex depth2ex
  Selection of some low-noise BJT amplifiers.}
\centering
\begin{tabular}{|c|cccc|c|c|}\hline
  & OP27\footnotemark[1]
         & LT1028\footnotemark[1] 
                & MAT02\footnotemark[2] 
                       & MAT03\footnotemark[2]
                              & \parbox{12ex}{unit} 
                                     & \parbox{12ex}{{\footnotesize MAT03}\\
                                       measured\footnotemark[3]%
                                \vrule width0pt height0ex depth0.5ex} 
                                       \\\hline
\vrule width0pt height2.5ex depth0ex
WHITE NOISE&&&&&&\\
noise voltage\footnotemark[4] $\sqrt{h_{0,v}}$     
  &   3  & 0.9  &  0.9 &  0.7 & \unit{nV/\sqrt{Hz}} &  0.8 \\
noise current\footnotemark[4] $\sqrt{h_{0,i}}$     
  &  0.4 &   1  &  0.9 &  1.4~\footnotemark[5]& \unit{pA/\sqrt{Hz}} & 1.2\\
noise power $2\sqrt{h_{0,v}h_{0,i}}$
  &$2.4{\times}10^{-21}$ 
         &$1.8{\times}10^{-21}$
                &$1.6{\times}10^{-21}$ 
                       & $2.0{\times}10^{-21}$
                              & \unit{W/Hz}
                                     &  $1.9{\times}10^{-21}$\\
noise temperature $T_w$
  &  174 &  130 &  117 &  142 &   K  & 139 \\
optimum resistance $R_{b,w}$
  & 7500 &  900 & 1000 &  500 & \ohm & 667 \\  
$2{\times}50$\ohm-input noise
  &  3.3 & 1.55 & 1.55 &  1.5 & \unit{nV/\sqrt{Hz}} & 1.5~%
                                                 \footnotemark[6]\\\hline
\vrule width0pt height2.5ex depth0ex
FLICKER NOISE&&&&&&\\
noise voltage\footnotemark[4] $\sqrt{h_{-1,v}}$    
  &  4.3 &  1.7 &  1.6 &  1.2 & \unit{nV/\sqrt{Hz}} & (~$0.4$~)%
                                                 \footnotemark[7] \\
noise current\footnotemark[4] $\sqrt{h_{-1,i}}$    
  &  4.7 &  16  &  1.6 &n.\,a.& \unit{pA/\sqrt{Hz}} &  11   \\
noise power $2\sqrt{h_{-1,v}h_{-1,i}}$
  & $4.1{\times}10^{-20}$ 
         & $5.3{\times}10^{-20}$
                & $5.1{\times}10^{-21}$ 
                       &  --  &  \unit{W/Hz}
                                     & (\ldots)\footnotemark[8] \\
1-Hz noise temperature $T_f$
  & 2950 & 3850 &  370 &  --  &   K  & (\ldots)\footnotemark[8] \\
optimum resistance $R_{b,f}$
  &  910 &  106 & 1000 &  --  & \ohm & (\ldots)\footnotemark[8] \\  
$2{\times}50$\ohm-input noise
  &  4.3 &  2.3 &  1.6 &  --  & \unit{nV/\sqrt{Hz}} &  1.1~%
                                                \footnotemark[6] \\\hline
\vrule width0pt height2.5ex depth0ex 
THERMAL DRIFT        
  &  200 &  250 &  100 &  300 & nV/K & --
\\\hline
\end{tabular}
\footnotetext[1]{Low-noise operational amplifier.}
\footnotetext[2]{Matched-transistor pair. 
  MAT02 is \textsc{npn}, MAT03 is \textsc{pnp}.
  Data refer to the pair, biased at $I_C=1$ mA.}
\footnotetext[3]{Some MAT03 samples measured in our laboratory.
  See Sec.~\protect\ref{sec:frontend}}
\footnotetext[4]{Power-law model of the spectrum, voltage or current,  
  $S(f)=h_0+h_{-1}f^{-1}+h_{-2}f^{-2}+\ldots$}
\footnotetext[5]{Obtained from the total noise with 100 k\ohm\ 
  input resistance.}
\footnotetext[6]{Measured on the complete amplifier 
  (Sec.~\protect\ref{sec:results}), independently
  of the measurement of the above $S_v$ and $S_i$.}
\footnotetext[7]{Derives from the noise current through $r_{bb'}$. 
  See Sec.~\protect\ref{sec:results}.}
\footnotetext[8]{Can not be compared to other data because voltage 
  and current are correlated.  See Sec.~\protect\ref{sec:results}.}
\end{minipage}
\end{sideways}
\end{table}

Table~\ref{tab:opa} compares a selection of low-noise bipolar
amplifiers.  The first columns are based on the specifications
available on the web
sites~\cite{www.analog-devices,www.linear-technology}.  The right-hand
column derives from our measurements, discussed in
Secs.~\ref{sec:frontend} and \ref{sec:results}.  Noise is described in
terms of a pair of random sources, voltage and current, which are
assumed independent.  This refers to the Rothe-Dahlke
model~\cite{rothe56ire}.  Nonetheless, a correlation factor arises in
measurements, due to the distributed base resistance $r_{bb'}$.
Whether and how $r_{bb'}$ is accounted for in the specifications is
often unclear.  The noise spectra are approximated with the power law
$S(f)=\sum_{\alpha}h_\alpha f^\alpha$.  This model, commonly used in
the domain of time and frequency, fits to the observations and
provides simple rules of transformation of spectra into two-sample
(Allan) variance $\sigma_y(\tau)$.  This variance is an effective way
to describe the stability of a quantity $y$ as a function of the
measurement time $\tau$, avoiding the divergence problem of the
$f^\alpha$ processes in which $\alpha\le-1$.
References~\cite{rutman78pieee} and \cite{rubiola01im} provide the
background on this subject, and application to operational amplifiers.

The noise power spectrum $2\sqrt{h_vh_i}$ is the minimum noise of the
device, i.e., the noise that we expect when the input is connected to
a cold (0~K) resistor of value $R_b=\sqrt{h_{v}/h_{i}}$, still under
the assumption that voltage and current are uncorrelated.  When the
input resistance takes the optimum value $R_b$, voltage and current
contributions to noise are equal.  The optimum resistance is $R_{b,w}$
for white noise and $R_{b,f}$ for flicker.  Denoting by $f_{c}$ the
corner frequency at which flicker noise is equal to white noise, thus
$f_{c,v}$ for voltage and $f_{c,i}$ for current, it holds that
$R_{b,w}/R_{b,f}=\sqrt{f_{c,i}/f_{c,v}}$.  Interestingly, with most
bipolar operational amplifiers we find
$f_{c,i}/f_{c,v}\approx50{-}80$, hence $R_{b,w}/R_{b,f}\approx7{-9}$.
Whereas we have no explanation for this result, the lower value of the
flicker optimum resistance is a fortunate outcome.  The equivalent
temperature is the noise power spectrum divided by the Boltzmann
constant $k=1.38{\times}10^{-23}$ J/K\@.  A crucial parameter of
Table~\ref{tab:opa} is the total noise when each input is connected to
a 50~\ohm\ resistor at room temperature.  This calculated value
includes noise voltage and current, and the thermal noise of the two
resistors.  In a complete amplifier two resistors are needed, at the
input and in the feedback circuit.

Still from Table~\ref{tab:opa}, the transistor pairs show lower noise
than the operational amplifiers, although the PNP pair is only
partially documented.  Experience indicates that PNP transistors are
not as good as NPN ones to most extents, but exhibit lower noise.  In
other domains, frequency multipliers and radio-frequency oscillators
make use of PNP transistors for critical application because of the
lower flicker noise.  Encouraged by this fact, we tried a differential
amplifier design based on the MAT03, after independent measurement of
some samples.

\section{Input Stage}\label{sec:frontend}
%
\begin{figure}[t]
  \centering\includegraphics[scale=1]{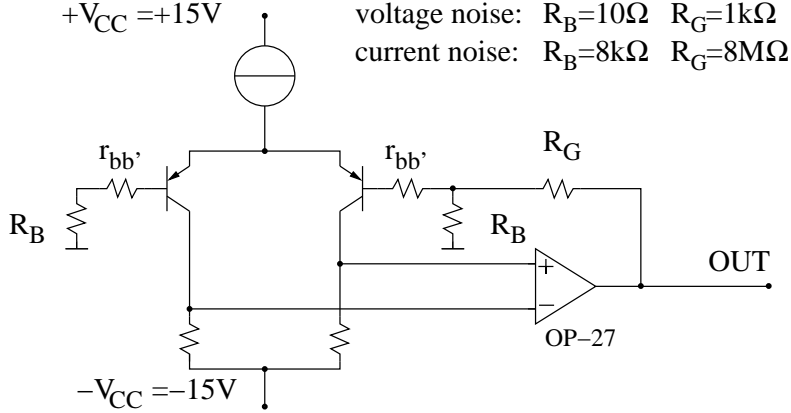}
\caption{Noise measurement of a transistor pair.  For clarity,
  the distributed base resistance $r_{bb'}$ is extracted from the
  transistors.}
\label{fig:measure-mat}
\end{figure}
The typical noise spectrum of the MAT03, reported in the data sheet,
shows an anomalous slope at low frequencies (0.1--1 Hz), significantly
different from $f^{-1}$.  This is particularly visible at low
collector current (10--100 $\mu$A), but also noticeable at $I_C=1$
mA\@.  We suspect that the typical spectrum reflects the temperature
fluctuation of the environment through the temperature coefficient of
the offset voltage $V_{OS}$ rather than providing information on the
flicker noise inherent in the transistor pair.  The measurement of a
spectrum from 0.1 Hz takes some 5 min.  At that time scale, in a
normal laboratory environment the dominant fluctuation is a drift.  If
the drift is linear, $v(t)=ct$ starting at $t=0$, the Fourier
transform is $V(\omega)=j\pi c\delta(\omega)-c/\omega^2$.  Dropping
off the term $\delta(\omega)$, which is a dc term not visible in a
log-log scale, the power spectrum density, i.e., the squared Fourier
transform, is
\begin{equation}
  \label{eq:f-drift}
S_v(\omega)=\frac{c^2}{\omega^4}  \qquad\mbox{or}\qquad
S_v(f)=\frac{(2\pi)^4c^2}{f^4}~~.
\end{equation}
A parabolic drift---seldom encountered in practice---has a spectrum
proportional to $f^{-6}$, while a smoothly walking drift tends to be
of the $f^{-5}$ type.  As a consequence, a thermal drift can be
mistaken for a random process of slope $f^{4}$ to $f^{5}$, which may
hide the inherent $f^{-1}$ noise of the device.  For this reason, the
test circuit (Fig.~\ref{fig:measure-mat}) must be enclosed in an
appropriate environment.  We used, with similar results, a Dewar flask
coupled to the environment via a heat exchanger, and a metal box
mounted on a heat sink that has a mass of 1 kg and a thermal
resistance of 0.6 K/W\@.  These odd layouts provide passive
temperature stabilization through a time constant and by eliminating
convection, and evacuate the small amount of heat (200 mW) dissipated
by the circuit.

\begin{figure}[t]
  \centering\includegraphics[scale=0.8,angle=0]{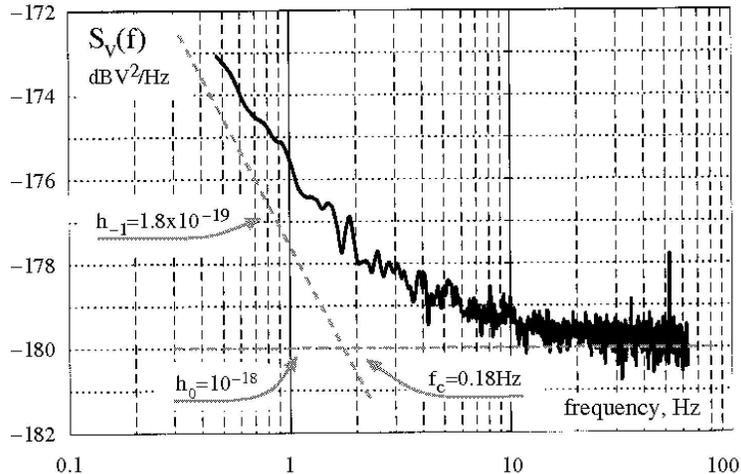}
\caption{Typical spectrum of the noise voltage.}
\label{fig:f695}
\end{figure}

Due to the low value of $r_{bb'}$ (15--20 \ohm) the current
measurement can be made independent of voltage noise, but not vice
versa.  Thus, we first measure the noise current setting
$R_B=8$~k\ohm, which is limited by the offset current; then we measure
the noise voltage setting $R_B=10$~\ohm.  A technical difficulty is
that at 1 Hz and below most spectrum analyzers---including our
one---must be coupled in dc, hence high offset stability is needed in
order to prevent saturation of the analyzer.  The measured spectra are
$S_i(f)=1.45{\times}10^{-24}+1.2{\times}10^{-22}f^{-1}$ \unit{A^2/Hz}
(i.e., 1.2\unit{pA/\sqrt{Hz}} white, and 11\unit{pA/\sqrt{Hz}}
flicker), and $S_v(f)=10^{-18}+1.8{\times}10^{-19}f^{-1}$
\unit{V^2/Hz} (i.e., 1\unit{nV/\sqrt{Hz}} white, and
425\unit{pV/\sqrt{Hz}} flicker).  The current spectrum is the inherent
noise current of the differential pair.  Conversely, with the voltage
spectrum (Fig.~\ref{fig:f695}) we must account for the effect of $R_B$
and $r_{bb'}$.  With our test circuit, the expected white noise is
$h_{0,v}=4kTR+2qI_BR\simeq1.7{\times}10^{-20}R$ \unit{V^2/Hz}, which
is the sum of thermal noise and the shot noise of the base current
$I_B$.  $R=2(R_B+r_{bb'})$ is the equivalent base resistance, while
the shot noise of the collector current is neglected.  Assuming
$r_{bb'}=16$~\ohm\ (from the data sheet), the estimated noise is
$h_{0,v}\simeq9{\times}10^{19}$ \unit{V^2/Hz}.  This is in agreement
with the measured value of $10^{-18}$ \unit{V^2/Hz}.  Then, we observe
the effect of the current flickering on the test circuit is
$R^2h_{-1,i}\simeq1.6{\times}10^{-19}$ \unit{V^2/Hz}.  The latter is
close to the measured value $1.8{\times}10^{-19}$ \unit{V^2/Hz}.
Hence, the observed voltage flickering derives from the current noise
through the external resistors $R_B$ and the internal distributed
resistance $r_{bb'}$ of the transistors.  Voltage and current are
therefore highly correlated.  As a further consequence, the product
$2\sqrt{h_{-1,v}h_{-1,i}}$ is not the minimum noise power, and the
ratio $\sqrt{h_{-1,v}/h_{-1,i}}$ is not the optimum resistance.  The
corresponding places in Table~\ref{tab:opa} are left blank.  Due to
the measurement uncertainty, we can only state that a true independent
voltage flickering, if any, is not greater than $4{\times}10^{-20}$
\unit{A^2/Hz}.  The same uncertainty affects the optimum resistance
$R_{b,f}$, which is close to zero.

The measured white noise is in agreement with the data sheet.  On the
other hand, our measurements of flicker noise are made in such unusual
conditions that the results should not be considered in contradiction
with the specifications, as the specifications reflect the the
low-frequency behavior of the device in a normal environment.

\section{Implementation and Results}\label{sec:results}

\begin{figure}[t]
\centering\includegraphics[scale=1]{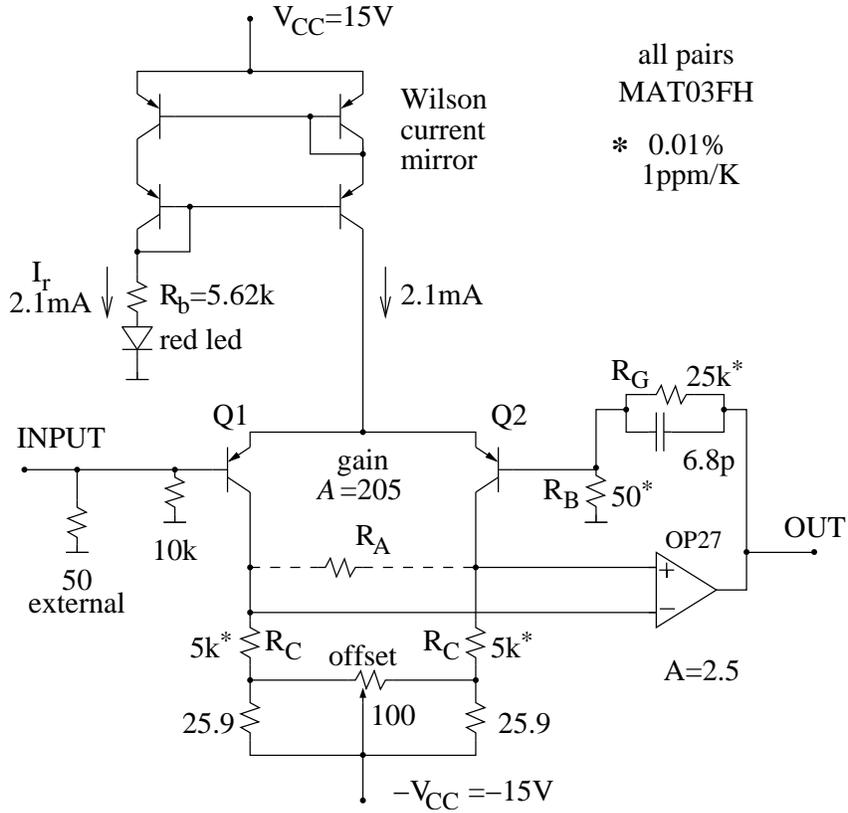}
\caption{Scheme of the low-noise amplifier.}
\label{fig:scheme}
\end{figure}

Figure~\ref{fig:scheme} shows the scheme of the complete amplifier,
inspired to the ``super low-noise amplifier'' proposed in Fig.~3a of
the MAT03 data sheet.  The NPN version is also discussed in
Ref.~\cite{franco:opa} (p.~344).  The original circuit makes use of
three differential pairs connected in parallel, as it is designed for
the lowest white noise with low impedance sources ($\ll50$~\ohm), like
coil microphones.  In our case, using more than one differential pair
would increase the flicker because of current noise.

The collector current $I_C=1.05$ mA results as a trade-off between
white noise, which is lower at high $I_C$, dc stability, which is
better at low dissipated power, flicker, and practical convenience.
The gain of the differential pair is $g_mR_C=205$, where
$g_m=I_C/V_T=41$~mA/V is the transistor transconductance, and $R_C=5$
k\ohm\ is the collector resistance.  The overall gain is
$1+R_G/R_B\simeq500$.  Hence the gain of the OP27 is of 2.5, which
guarantees the closed-loop stability (here, oscillation-free
operation).  If a lower gain is needed, the gain of the differential
stage must be lowered by inserting $R_A$.  The trick is that the
midpoint of $R_A$ is a ground for the dynamic signal, hence the
equivalent collector resistance that sets the gain is $R_C$ in
parallel to $\frac{1}{2}R_G$.  The bias current source is a cascode
Wilson scheme, which includes a light emitting diode (LED) that
provides some temperature compensation.

The stability of the collector resistors $R_C$ is a crucial point
because the voltage across them is of 5~V\@.  If each of these
resistors has a temperature coefficient of $10^{-6}$/K, in the worst
case there results a temperature coefficient of 10 $\mu$V/K at the
differential output, which is equivalent to an input thermal drift of
50~nV/K\@.  This is 1/6 of the thermal coefficient of the differential
pair.  In addition, absolute accuracy is important in order to match
the collector currents.  This is necessary to take the full benefit
from the symmetry of the transistor pair.

\begin{figure}[t]
\centering\includegraphics[width=\textwidth]{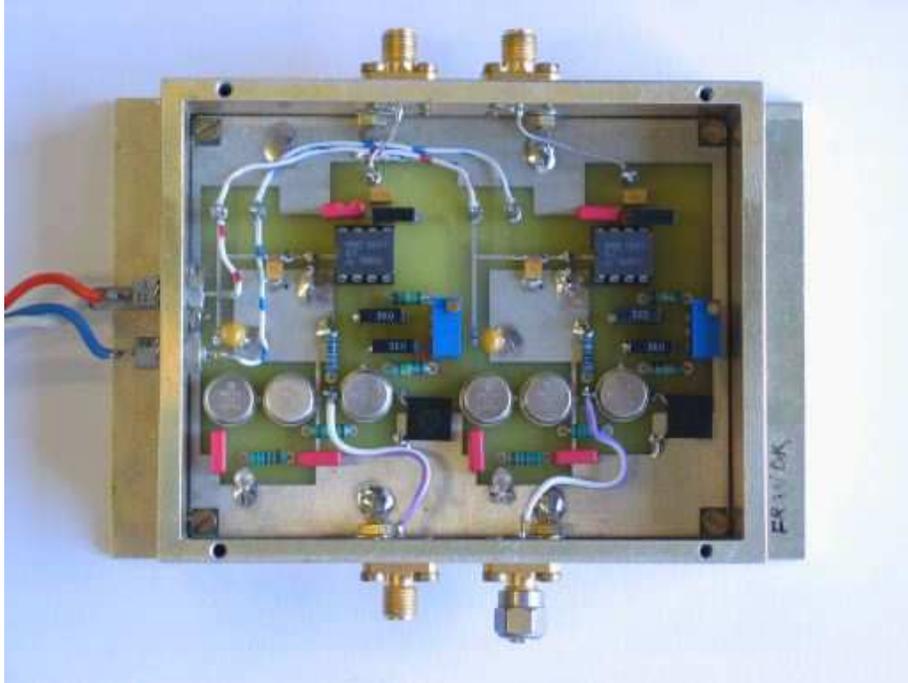}
\caption{Prototype of the low-noise amplifier.}
\label{fig:prototype}
\end{figure}

Two equal amplifiers are assembled on a printed circuit board, and
inserted in a $10{\times}10{\times}2.8$ \unit{cm^3}, 4 mm thick
aluminum box (Fig.~\ref{fig:prototype}).  The box provides thermal coupling to the environment
with a suitable time constant, and prevents fluctuations due to
convection.  $LC$ filters, of the type commonly used in HF/VHF
circuits, are inserted in series to the power supply, in addition to
the usual bypass capacitors.  For best stability, and also for
mechanical compatibility with our equipment, input and output
connector are of the SMA type.  Input cables should not PTFE-insulated
because of piezoelectricity (see the review
paper~\cite{fukada00uffc}).

\begin{figure}[t]
  \centering\includegraphics[scale=0.8]{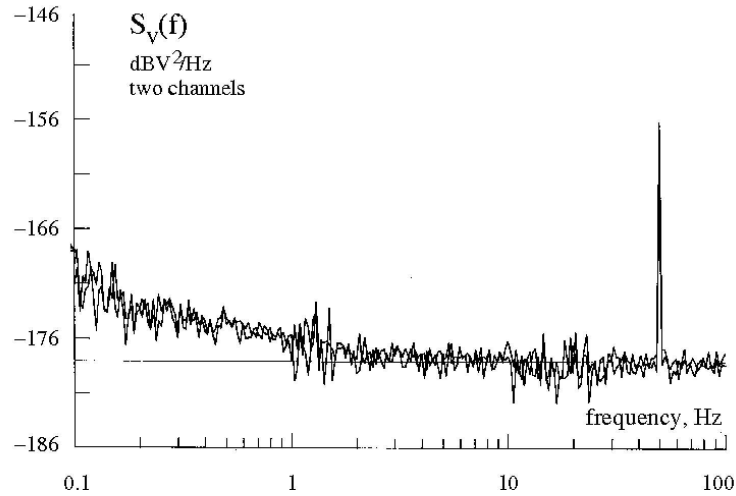}
\caption{Residual noise of the complete amplifier, 
  input terminated to a 50~\ohm\ resistor.}
\label{fig:f691}
\end{figure}

Figure~\ref{fig:f691} shows the noise spectrum of one prototype input
terminated to a 50~\ohm\ resistor.  The measured noise is
$\sqrt{h_0}=1.5$ \unit{nV/\sqrt{Hz}} (white) and $\sqrt{h_{-1}}=1.1$
\unit{nV/\sqrt{Hz}} (flicker).  The corner frequency at which the
white and flicker noise are equal is $f_c=0.5$ Hz.  Converting the
flicker noise into two-sample (Allan) deviation, we get
$\sigma_v(\tau)=1.3$ nV, independent of the measurement time $\tau$.

\begin{figure}[t]
  \centering\includegraphics[scale=0.8]{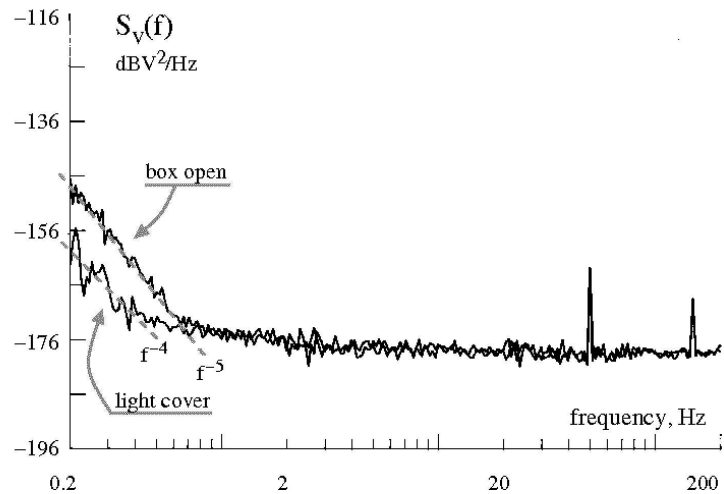}
\caption{Thermal effects on the amplifier.}
\label{fig:f694}
\end{figure}

Finally, we made a simple experiment aimed to explain in practical
terms the importance of a proper mechanical assembly.  We first
removed the Al cover, exposing the circuit to the air flow of the
room, yet in a quiet environment, far from doors, fans, etc., and then
we replaced the cover with a sheet of plain paper (80 \unit{g/m^2}).
The low-frequency spectrum (Fig.~\ref{fig:f694}) is
$5{\times}10^{-19}f^{-5}$ \unit{V^2/Hz} in the first case, and about
$1.6{\times}10^{-19}f^{-4}$ \unit{V^2/Hz} in the second case.  This
indicates the presence of an irregular drift, smoothed by the paper
protection.  Interestingly, Hashiguchi~\cite{sikula03arw} reports
on thermal effects with the same slope and similar cutoff frequencies,
observed on a low-noise JFET amplifier for high impedance sources.

\def\bibfile#1{/Users/rubiola/Documents?workocs/bib/#1}
\bibliographystyle{amsalpha}
\bibliography{\bibfile{ref-short},%
              \bibfile{references},%
              \bibfile{rubiola}}

\end{document}